\documentclass{jfm}
\usepackage{graphicx}
\usepackage{epstopdf, epsfig}
\usepackage{comment}
\usepackage{bm}
\usepackage[normalem]{ulem}
\usepackage[shortlabels]{enumitem}
\usepackage{CJK}
\usepackage{amssymb,euscript,latexsym,amsmath}
\usepackage{graphicx}
\usepackage{color}
\usepackage{epstopdf}
\usepackage{setspace}
\usepackage{subfigure}
\usepackage[]{natbib} % for other bibliography things
\usepackage{chngpage}
\usepackage{textgreek}
\usepackage{tikz}
\def\h{{\rm h}}
\def\p{{\rm p}}
\def\f{{\rm f}}
\def\b{{\rm b}}
\def\w{{\rm w}}
\def\ii{{\rm i}}

\shorttitle{Flow-mediated schooling of heaving and pitching swimmers}
\shortauthor{S. Heydari and E. Kanso}

\title{Passively stable patterns in the swimming of heaving and pitching plates}
\title{School emergent properties depend on the swimmers' flapping mode}
\title{School cohesion, speed, and efficiency are modulated by the swimmers flapping motion}
\author{Sina Heydari
 \and Eva Kanso
  \corresp{\email{kanso@usc.edu}}}

\affiliation{Aerospace and Mechanical Engineering,  University of Southern California, Los Angeles, California 90089, USA}

\begin{document}

\maketitle

\begin{abstract}
Fish schools are ubiquitous in marine life. Although flow interactions are thought to be beneficial for schooling, their exact effects on the speed, energetics, and stability of the group remain elusive. {Recent numerical simulations and experimental models suggest that flow interactions stabilize in-tandem formations of flapping foils.} Here, we employ a minimal vortex sheet model that captures salient features of the flow interactions among flapping swimmers, and we study the free swimming of a pair of in-line swimmers driven with identical heaving or pitching motions. We find that, independent of the flapping mode, heaving or pitching, the follower passively stabilizes at discrete locations in the wake of the leader, consistent with the heaving foil experiments, but pitching swimmers exhibit tighter and more cohesive formations.
Further, in comparison to swimming alone, pitching motions increase the energetic efficiency of the group while heaving motions result in a slight increase in the swimming speed.  
{A deeper analysis of the wake of a single swimmer sheds light on the hydrodynamic mechanisms underlying pairwise formations.} These results recapitulate that flow interactions provide a passive mechanism that promotes school cohesion, and 
afford novel insights into the role of the flapping mode in controlling the emergent properties of the school.
\end{abstract}


\section{Introduction}

Fish schools are ubiquitous in aquatic life, with half of the known fish species thought to exhibit schooling behavior during some phase of their life cycle~\citep{Shaw1978}. However, the role of the fluid medium as a mediator of the physical interactions between swimming fish remains unclear~\citep{Partridge1979, Partridge1982}.
Experimental evidence suggests that fish modify their motions and reduce muscular effort when swimming in vortex-laden flows~\citep{Liao2003}.
These findings support a long-standing but controversial hypothesis that schooling provides hydrodynamic benefits as fish move within the flows generated by others~\citep{Weihs1973, Weihs1975,Abrahams1985, Liao2007}. 
A direct assessment of this hypothesis in biological and physical models remains a challenge because of the complexity in resolving the hydrodynamics of unsteady swimming at high Reynolds numbers in single~\citep{Wolfgang1999,Triantafyllou2000,Borazjani2008} and multiple interacting swimmers~\citep{Liao2007,Gazzola2016, Verma2018}. 
Simplifications based on crystalline school arrangements and ideal flow models indicate that fish within a planar formation, with diamond-shaped unit cell, benefit energetically from near-field interactions with the wakes of upstream neighbors~\citep{Weihs1973}, whereas far-field interactions serve to passively stabilize the formation~\citep{Tsang2013}. 
These crystal lattice models do not capture that fish exhibit variable arrangements in field and laboratory experiments~\citep{Partridge1979,Marras2015},
%Critique notwithstanding, recent experiments demonstrate that fish modify their motions and reduce muscular effort when swimming in vortex-laden flows \citep{Liao2003}, 
and the broader question of how flow interactions benefit schooling remains unresolved. 

{Physical models and numerical simulations of mechanically actuated foils found that, at the single swimmer level, flapping foils share with their biological counterparts many common aspects of the flows, forces, and energetics~\citep{Blondeaux2005,Dong2006, Buchholz2008,Dabiri2009,Lauder2011, Wen2013}. 
A key similarity is the reverse von K\'{a}rm\'{a}n wake left by both flapping foils and fish
%: a staggered array of counter-rotating vortices surrounding a sinuous jet-like flow
~\citep{Taneda1965,Triantafyllou1993}. 
Subsequently, several numerical and experimental studies used pairs of flapping foils to understand multi-swimmer interactions.  \cite{Zhu2014} were first to examine,  in the context of the immersed boundary method, the effects of pairwise hydrodynamic interactions on the self-propulsion of flapping flexible swimmers in tandem configuration. 
% The flexible filaments were driven into heaving motion at the leading edge. 
Flow-mediated interactions were found to stabilize the swimmers in particular spacings and to reduce the energetics cost of swimming in the follower. 
Experimental studies on heaving rigid foils confined to in-line positions and freely swimming in tandem were also found to assume one of several particular spacings, stabilized by the flow interactions~\citep{Becker2015, Ramananarivo2016, Newbolt2019}. These observations have since been confirmed in several numerical studies \citep{Park2018, Peng2018,Dai2018, Lin2020}.}
 Here, we investigate the speed, energetics, and stability of these planar formations using a mathematical model of self-propelling and interacting swimmers that flap by either heaving or pitching. 
%We address, in the context of the model, the range and strength of hydrodynamic interactions between a pair of free swimmers and we analyze the effect of these interactions on the speed, stability, and energetic cost of the emergent formations. 

%----
\begin{figure*}
\centering
\includegraphics[scale=1]{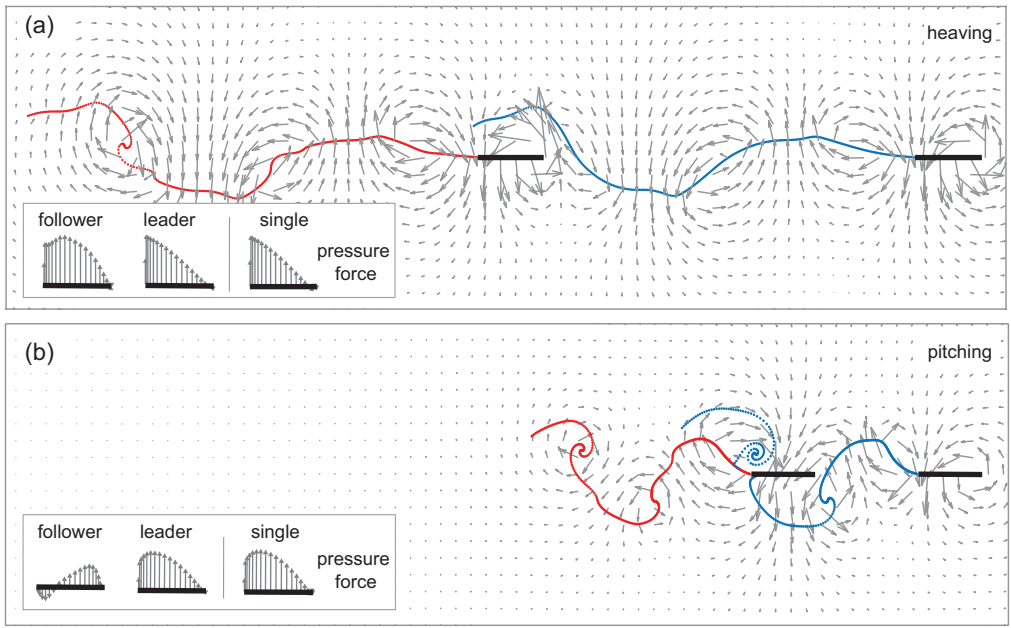}
\caption[]{\footnotesize
A pair of swimmers undergoing (a) heaving motions at amplitude $A_\h = 0.3$ and (b) pitching motions at amplitude $A_\p = 15^{\circ}$. Snapshots of the velocity field (grey arrows) and free vortex sheet of the leader (blue) and follower (red) are taken after steady-state swimming is reached at a time instant where both swimmers are flapping downwards.
% at $t=120.5$ flapping periods for heaving and $t=40$ flapping periods for pitching. 
Insets depict the pressure forces acting on each swimmer in the pairwise formation in comparison to 
a single swimmer undergoing the same prescribed motion.}
\label{fig:flow_field}
\end{figure*}
%------

Existing mathematical models of flow interactions in fish schools vary in the degree of fidelity to the fluid dynamics and sensory-feedback control at the swimmer level.
Ideal flow models -- based on a dipolar far-field approximation \citep{Tchieu2012} -- with no feedback control have been used to assess the effect of passive flow interactions on the stability of pairwise \citep{Kanso2014,Kanso2015} and diamond lattice formations \citep{Tsang2013} and the advantages of flapping out-of-phase \citep{Kanso2009}. This far-field flow model coupled to visual feedback control, either in the form of behavioral rules~\citep{Filella2018} or learning algorithms~\citep{Gazzola2016}, was used to analyze the fish collective dynamics. Fish were shown to exhibit a novel collective turning mode and to swim faster thanks to the fluid~\citep{Filella2018}. 
Near-field fish-wake interactions were also accounted for in
ideal flow models with no feedback control, such as the vortex street model used by \cite{Weihs1973} or the phenomenological model derived in~\cite{Oza2019} to assess the efficiency of lattice formations. 
High-fidelity computational fluid dynamics coupled to reinforcement learning algorithms were recently  implemented in pairwise interactions to optimize the flapping motion of the follower fish for harnessing the wake of the leader \citep{Verma2018}. 

In this paper, we analyze pairwise interactions of heaving and pitching swimmers in the context of the vortex sheet model (see Figure~\ref{fig:flow_field}). The vortex sheet model has been used extensively to analyze problems of  fluid-structure interactions, including ring formation at the edge of  a circular tube~\citep{Nitsche1994} and wakes of oscillating plates~\citep{Jones2003,Sheng2012}, falling cards~\citep{Jones2005},  flapping 
flexible flags~\citep{Alben2008,Alben2009}, swimming plates~\citep{Wu1971} and hovering flyers~\citep{Huang2016,Huang2018}. Here, we use the implementation of \citep{Nitsche1994}. We specifically focus on the effect of streamwise flow interactions on the swimming motion of heaving and pitching plates, and find that ordered formations emerge spontaneously via these interactions, independent of the flapping mode, consistent with heaving foil experiments \citep{Ramananarivo2016, Newbolt2019} and {numerical simulations \citep{Zhu2014,Park2018, Peng2018,Lin2020}}. However, the flapping mode, heaving or pitching, affects the speed and energetics of these formations as well as their robustness to streamwise perturbations. {We describe a specific hydrodynamic mechanism that explains the energetic and stability differences associated with each flapping mode. }

\section{Problem formulation}
\label{sec:problem}

A swimmer is modeled as a rigid plate of length $2l$, small thickness $e\ll l$, and homogenous density $\rho$, submerged in an unbounded, planar, fluid domain of density $\rho_f$. The swimmer's mass per unit depth is given by $m = 2 \rho e l$. An inertial frame $(\mathbf{e}_x,\mathbf{e}_y,\mathbf{e}_z)$ is introduced, such that $(\mathbf{e}_x,\mathbf{e}_y)$ span the plane of motion. The vector $\mathbf{x} \equiv (x,y)$ denotes the position of the leading edge of the swimmer in the $(\mathbf{e}_x,\mathbf{e}_y)$ plane, and the angle $\theta$ its orientation relative to the $\mathbf{e}_x$-direction (see Appendix~\ref{sec:appendix} and Figure~\ref{fig:appendix})

The swimmer is free to move in the $\mathbf{e}_x$-direction under periodic heaving or pitching motions. Heaving consists of periodic lateral motions in the $y$-direction, of amplitude $A_\h$, at fixed angle $\theta = 0$. Pitching refers to angular oscillations $\theta$ of amplitude $A_\p$, with zero lateral motion $y=0$ at the leading edge. The frequency of these heaving and pitching motions is denoted by $f$. Hereafter, we scale all parameter values using $l$ as the characteristic length scale, $1/f$ as the characteristic time scale, and $\rho_f l^2$ as the characteristic mass per unit depth. Accordingly, velocities are scaled by $lf$, forces by $\rho_f f^2 l^3$, moments by $\rho_f f^2 l^4$, and power by $\rho_f f^3 l^4$.

In dimensionless form, the heaving and pitching motions are given by
%---
\begin{equation}
\begin{split}
\textrm{Heaving:} & \quad  y(t) = A_\h \sin(2 \pi t), \quad \theta(t) = 0, \\ \textrm{Pitching:} & \quad \theta(t)  = A_\p \sin(2\pi t),\quad y(t) = 0 .
\end{split}
\end{equation}
%---
The equation of motion governing the free swimming $x(t)$ is given by Newton's second law
% Note that the two plates undergo the same motion in the $y$-direction and $\theta$-direction, but are not constrained in any way in the $x$-direction. Namely, they are free to move in this direction as 
%The equation of motion for each plate in the $x$-direction is given by the balance of linear momentum on that plate, namely,
%--
{
\begin{equation}
%m \ddot{x}= \sum F_{x}  =  F_{x}^{\rm le} + F_{x}^{\rm hp} + F_{x}^{\rm sd} .
m \ddot{x}=  - F \sin\theta + S \cos\theta - D \cos\theta.
\label{eq:eom}
\end{equation}
%--
Here, the hydrodynamic forces acting on the swimmer consist of a leading edge suction force $S$, a pressure force $F$ acting in the direction normal to the swimmer, and a skin drag force $D$ acting tangentially to the swimmer in the opposite direction to its motion.} The drag force $D$ is introduced to emulate the effect of fluid viscosity, while the hydrodynamic pressure force $F$ is calculated in the context of the inviscid vortex sheet model.  A detailed description of the method and its numerical implementation can be found in \cite{Nitsche1994,Huang2018}, and a brief overview is given in Appendix \ref{sec:appendix}. Detailed expressions of the fluid forces and moments acting on the swimmer are given in~\ref{app:forces}.
%These forces include all the hydrodynamic forces from the fluid and the interaction between different swimmers. 

To assess the swimming performance, we use four metrics: the period-averaged swimming speed $U =  \int_t^{t+1} \dot{x} dt$ at steady state, the thrust force $T = S \cos \theta - F \sin \theta$, the input power $P$ required to maintain the prescribed heaving or pitching motions (see details in Appendix~\ref{sec:efficiency}), and the cost of transport defined as the input power $P$ divided by the swimming speed $U$.

%We consider two swimmers placed in tandem as shown in Figure~\ref{fig:flow_field}, in which case.

%----
\begin{figure*}
\centering
\includegraphics[scale=1]{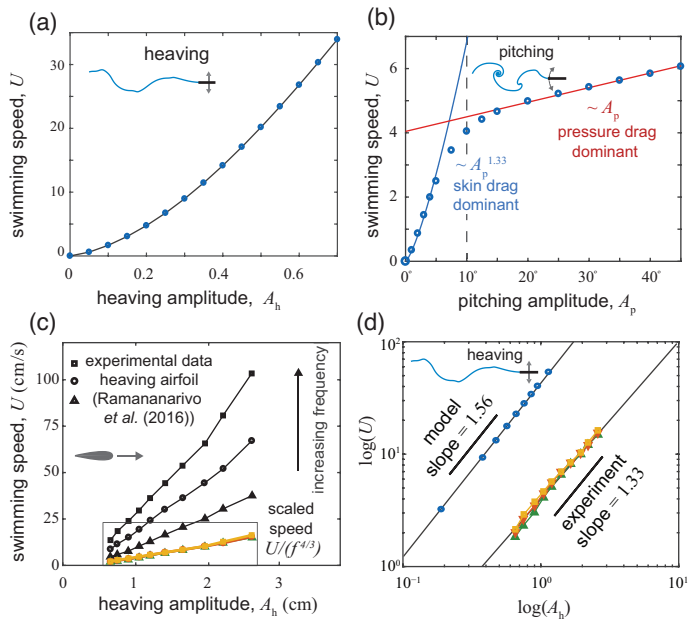}
\caption[]{\footnotesize Swimming speed versus flapping amplitude for single swimmers. (a) Average swimming speed at steady state for a heaving swimmer. (b) Average swimming speed at steady state for a pitching swimmer. At small $A_\p$, skin drag is dominant and the speed scales super-linearly with $A_\p$. For $A_\p>10^o$, pressure drag is dominant and speed scales linearly with $A_\p$.  (c) Experimental data (black markers) of average swimming speed of a heaving foil  \cite[Figure 2]{Ramananarivo2016}; the data collapses when scaled by the heaving frequency $f^{4/3}$ (yellow markers). (d) Comparing the swimming speed of our heaving swimmer model (blue circles) to the frequency-scaled experimental data shown in (a), on a log-log scale. Both model and experimental results scale super-linearly with heaving amplitude.  
%(a) The black markers denote the experimentally measured swimming speed as a function of heaving amplitude for different frequency values. The colored markers are the measured data normalized by the heaving frequency according to our scaling law Eq.~\ref{eq:scale_heaving}. The collapse of the three sets of data to a single curve confirms the derived scaling law. (b) The average swimming speed from the model (blue dots) and the experimental data (colored markers) for a single heaving swimmer versus amplitude, plotted on a log-log scale. (c) The blue dots represent the average swimming speed from the model of a single pitching swimmer and the two lines represent the scaling laws fitted to the data according to Eq.~\ref{eq:scale_pitching}.
}
\label{fig:compare}
\end{figure*}
%----------

\section{Single swimmers: numerical results and scaling analysis}

We solve~\eqref{eq:eom} in the case of a single swimmer and compute the period-average swimming speed at steady state. In Figure~\ref{fig:compare}(a) and (b), we show the steady state speed for heaving and pitching swimmers, respectively, as a function of the flapping amplitude. In both cases, the speed increases monotonically, albeit that, when pitching, the increase scales differently at small amplitudes.
To get insight into how the swimming speed $U$ scales with the heaving and pitching amplitudes and frequency, it is instructive to use a simple scaling analysis.
% to further understand the super-linear relationship between the swimming speed of a single swimmer and the heaving amplitude. 
%We choose this particular swimming property because corresponding experimental data are provided in \cite{Ramananarivo2016}.

{At steady state, the sum of forces acting on the swimmer is zero on average.  For heaving swimmers, the dominant forces are those due to leading edge suction and viscous skin drag \citep{Garrick1937}. In dimensional form, the suction force scales as $\rho_\f (2l)C_s^2U^2 $ , where the coefficient $C_s$ scales linearly with the effective angle of attack. In a heaving flat plate the effective angle of attack is given by $ \dot{y}/ U \sim A_\h f  / U$
 \citep{Garrick1937,Floryan2017,Franck2017,Smits2019}.  As a result, the suction force scales as $\rho_\f (2l)(A_\h f)^2$. 
%the inertial added mass and viscous skin drag.  In dimensional form, the inertial force scales as $\rho_\f (2l)(A_\h f)^2$, where $\rho_\f (2l)A_\h$ represents the mass of the  laterally displaced fluid and $A_\h f^2$ its acceleration. 
Skin drag scales as $\rho_f (2l) C_\f U^2$, where $C_\f \sim \sqrt{\mu/\rho_f (2l) U}$ is the drag coefficient based on adapting Blasius theory to this inviscid fluid model (see Appendix C and \cite{White1979}). Balancing suction and drag forces, we arrive at $(A_h f)^2 \sim U^{3/2}$, which leads to
%----
 \begin{equation}
 \label{eq:scale_heaving}
 \textrm{Heaving: } U\sim (A_\h f)^{4/3}.
 \end{equation}
%----
}
The swimming speed scales super-linearly with the heaving amplitude and frequency. We test this scaling law in light of the experimental results of~\cite[Figure 2]{Ramananarivo2016}. The black data points in Figure~\ref{fig:compare}(c) represent the experimentally measured swimming speed as a function of heaving amplitude. The different marker shapes represent three different heaving frequencies used in the experiments ($f=1, 2, 3$). We scaled the data by the heaving frequency according to our derived scaling law in \eqref{eq:scale_heaving}. The scaled data (colored markers) collapses on a single curve, indicating that our scaling analysis is sound. In Figure~\ref{fig:compare}(d), we plot, using a log-log scale, the swimming speed obtained from our model in Figure~\ref{fig:compare}(a) (blue dots) and experimental data (colored markers) versus the heaving amplitude. The slope of each line represents the power law that governs the relationship between the two quantities. In both the model and the experiment, the swimming speed depends super-linearly on the amplitude of heaving, however, the dependence is slightly stronger in the model.

The steady state speed of the pitching swimmer scales differently depending on the flapping amplitude because the dominant drag forces acting on the swimmer differ.
%because the dominant forces acting on the swimmer differ depending on the pitching amplitude $A_\p$.
At small pitching amplitude $A_\p$, the swimmer is almost parallel to the swimming direction, hence skin drag is dominant leading to the same scaling law as in the heaving case. At large amplitude $A_\p$, pressure drag is dominant; it is well known that pressure drag scales as $U^2$; see, e.g., \cite{Moored2019}. Balancing inertia and pressure drag, we arrive at $U \sim A_p f$. Put together, we have
%--
\begin{equation}
\label{eq:scale_pitching}
 \textrm{Pitching:} 
 \begin{cases} \textrm{small} \ A_\p:  U \sim (A_\h f)^{4/3}, \\ \textrm{large} \ A_\p \ : U \sim A_\p f.
 \end{cases}
\end{equation}
%--
These scaling laws fit remarkably well the numerical results in Figure~\ref{fig:compare}(b).

%%%%%%%%%%%%%%%%%%%%%%%%%%
\section{Pairwise formations: stability, speed, and energetics}
We examine the steady state behavior of a pair of swimmers undergoing heaving and pitching motions while freely interacting via the fluid medium.
In Figure~\ref{fig:flow_field}, we show snapshots of the flow field (grey arrows) and free vortex sheets in the case when the leader (blue) and follower (red) are heaving at  $A_\h = 0.3$ (Figure~\ref{fig:flow_field}a) and pitching at $A_\p = 15^{\circ}$ (Figure~\ref{fig:flow_field}b).
The snapshots are taken after the pair has reached steady state swimming in the positive $x$-direction, and passively locked into a constant separation distance. At these flapping amplitudes, the heaving swimmers experience longer transience and swim faster, whereas the pitching swimmers rapidly lock into a tighter formation (see Supplemental Movie 1).

An analysis of the hydrodynamic pressure forces $-F \sin\theta \mathbf{e}_x + F \cos\theta \mathbf{e}_y$, where $F$ is given in Appendix~\ref{app:forces}, acting on each swimmer shows that compared to a single swimmer, the distribution on the leader remains relatively unchanged. However, the force distribution on the follower is affected by the wake of the leader, and the effect is more pronounced for pitching swimmers; see insets in Figure~\ref{fig:flow_field}(a) and (b). Specifically in the pitching case, the follower experiences less resistance from the fluid, and a favorable force distribution (in the same direction of flapping) at the swimmer's tail. At the instant shown in Figure 1(b), the downward flow due to the vortex sheet created by the leader helps the follower in its downward pitching motion.

%----------
\begin{figure*}
\centering
\includegraphics[scale=1]{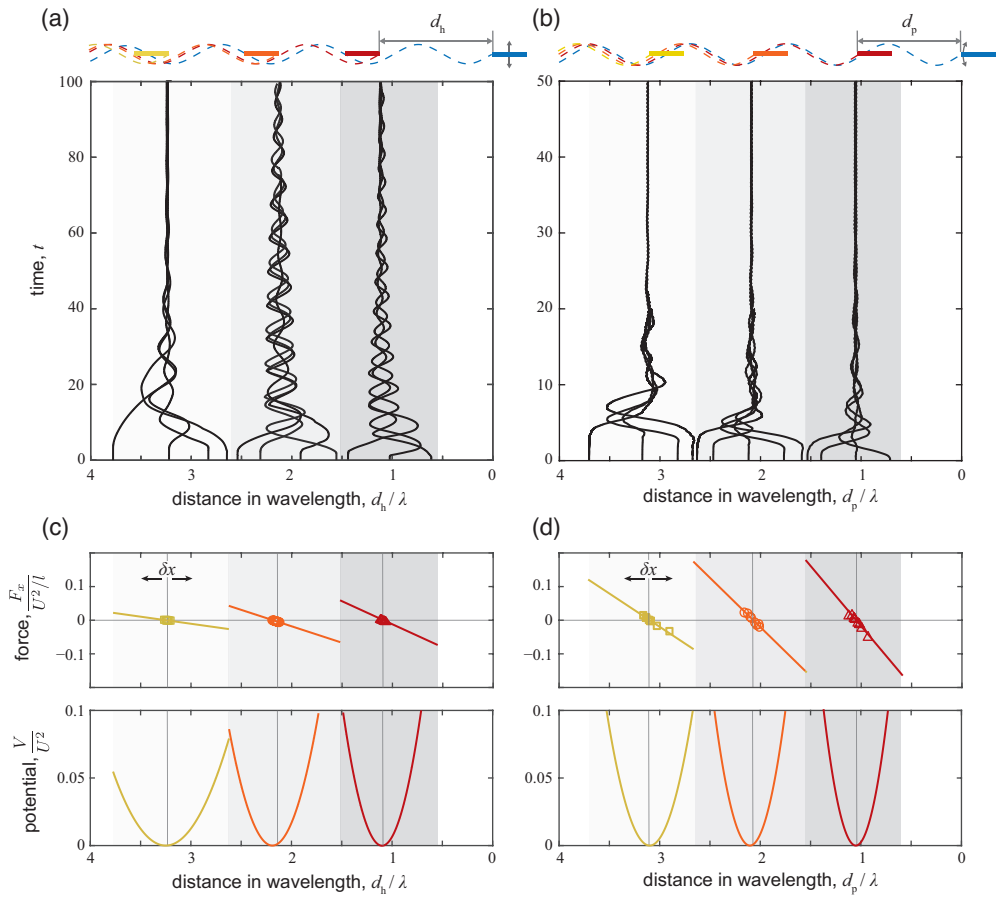}
\caption[]{\footnotesize 
 Emergence of passive stable formations in a pair of heaving swimmers ($A_\h = 0.3$) and of pitching swimmers ($A_\p = 15^{\circ}$). (a) For heaving swimmers, the follower stabilizes at one of many discrete positions behind the leader where the gap (tail-to-head) distance $d_\h$  is close to integer multiple of of the wavelength $\lambda = U/f$ of the leader motion. (b) For pitching swimmers, the follower stabilizes at locations such that the tail-to-tail distance $d_\p$ is close to integer multiples of $\lambda$.
 Basins of attraction of each the first three equilibria are depicted in gradually more faint shades of grey. %Starting from different initial distances, the normalized distances converge close to digitized values in heaving and pitching pair of swimmers.
% The pairwise formations of (c) heaving and (d) pitching swimmers are stable to perturbations.
(c) and (d) Linear stability analysis: we perturb the position of the follower about each of these equilibria and compute the total hydrodynamic force $F_x$. 
We simultaneously sample data from the change in  $F_x$ and perturbation strength $\delta x$, and plot $\delta F_x$ versus $\delta x$. Clearly, $\delta F_x$ acts as a restoring force. Taking the slope of $\delta F_x$, we construct the hydrodynamic potential $V$ on the follower. The potential well is deepest at the first equilibrium where the hydrodynamic interactions are strongest.
}
\label{fig:stability}
\end{figure*}
%----------

In Figure~\ref{fig:stability}, we vary the initial separation distance between the two swimmers for the examples shown in Figure~\ref{fig:flow_field}. We find that for both heaving and pitching, the follower tends to one of several discrete locations behind the leader at nearly digital values of $d_\h/\lambda$ and $d_\p/\lambda$, respectively, where
 $d_\h$ is the tail-to-head distance, $d_\p$ the tail-to-tail distance, and $\lambda = U/f$ the wavelength of the leader's swimming trajectory; see Figure~\ref{fig:stability}(top). 
Depending on initial conditions, the leader and follower reach one of these separation distances and swim together in ordered formation. 
{These findings are consistent with observations on heaving foils \citep{Zhu2014,Ramananarivo2016, Park2018, Peng2018,Lin2020}}. 
 %Note that the results in %Figures~\ref{fig:flow_field},~\ref{fig:example}(b), and \ref{fig:amplitude}(b) correspond to the first preferred position behind the leader with the shortest separation distance ($d_\h/\lambda = 1$ and $d_\p/\lambda = 1$) and strongest hydrodynamic interactions.

%The emergence of these discrete relative equilibria in nonlinear simulations indicates that they are  stable. 
We examine the nonlinear basins of attraction of these equilibria by varying the initial separation distance $d_\h$ and $d_\p$ between the two swimmers; The basin of attraction of each relative equilibrium is highlighted in a different shade of grey in Figure~\ref{fig:stability}(a,b). The pitching swimmers converge more rapidly to the corresponding equilibria, indicating that these equilibria are stronger attractors in pitching than in heaving. Further, the wavelength $\lambda = U/f$ is smaller  in pitching, and so is the actual separation distance at equilibria
($d_\p< d_\h$), indicating that pitching swimmers exhibit tighter formations.

To quantitatively assess the linear stability of these equilibria, we perturb the position of the follower about each equilibrium in the positive and negative $x$-direction with an initial perturbation of size $\delta x/l = 0.5$ and we calculate the corresponding change in $\delta x$ and change in the total hydrodynamic force $\delta F_x = \delta(- F \sin\theta + S \cos\theta - D \cos\theta)$  acting on the follower in the $x$-direction.
We scale the change in total force by $U^2/l$ and the perturbation from equilibrium by $d_\cdot/\lambda$, where $d_\cdot$ is either $d_\h$ or $d_\p$.
We sample simultaneously the scaled change in total force $\delta F_x$ and scaled perturbation strength $\delta x$ and we plot the results in the first row of Figure~\ref{fig:stability}(c,d). The results  are depicted in red $\triangle$ markers for the first stable position, and in orange $\circ$ and yellow $\Box$ markers for the second and third positions, respectively. Straight line fit for each of these data sets results in straight lines with negative slopes, implying that, for each of these equilibrium positions, the hydrodynamic force acts as a restoring force $\delta F_x = - K\delta x$ that keeps the formation stable. Here, $K$ is obtained numerically from the straight line fit. The value of $K$ depends monotonically on the equilibrium position of the follower, with highest value at the first equilibrium ($d_\h/\lambda \approx 1$ and $d_\p/\lambda \approx 1$). The first equilibrium is most stable because hydrodynamic interactions are strongest at closer distance. We write $\delta F_x = - \partial V/\partial(\delta x)$, where $V = K (\delta x)^2/2$ is the hydrodynamic potential function around the equilibrium $\delta x =0$. For both pitching and heaving, the formation is stable with weaker stability for larger inter-swimmer distance. In the pitching formation the potential well is deeper (by about $50\%$) for all equilibria, indicating faster convergence to the respective equilibrium.
  
 %----------
\begin{figure*}
\centering
\includegraphics[scale=1]{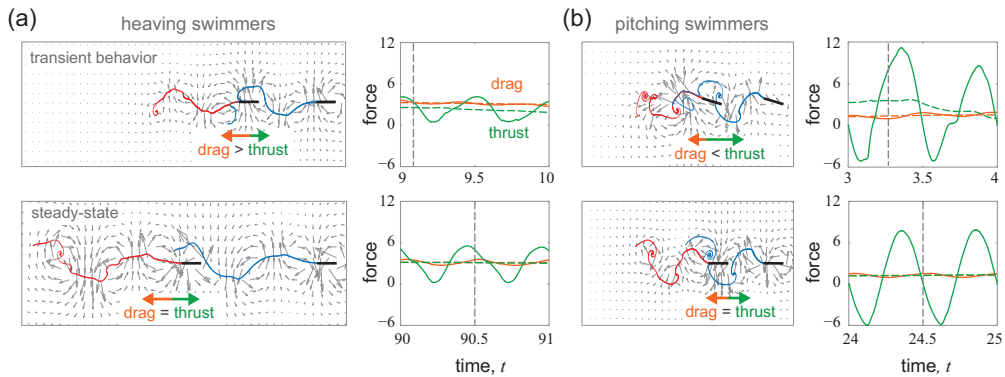}
\caption[]{\footnotesize {Hydrodynamic forces on the follower act as restoring forces. 
Snapshots of pairs of swimmers undergoing (a) heaving and (b) pitching motion during transient and steady-state formation. Green (thrust) and orange (drag) arrows represent period-averaged hydrodynamic forces acting on the follower. Right columns in (a) and (b) show the instantaneous thrust and skin drag (solid lines) and their period-averaged values (dashed lines) over one flapping period during transient and steady-state formation. The grey dashed lines denote the time instance of the snapshots shown to the left.
} 
}
\label{fig:transient}
\end{figure*}
%---------- 

{To further examine the hydrodynamic interactions mediating these equilibria, we plot snapshots of the free vortex sheets and flow field for the pair of heaving and pitching swimmers  in Figure~\ref{fig:transient}(a, b),
We report two instances taken during the transient and steady-state motion, and from each regime, we report the hydrodynamic thrust and skin drag over one period of flapping and their time-period average. 
% The green and orange arrows denote the magnitude of the period-averaged hydrodynamic forces acting on the follower at the specific instances shown.
When the follower gets too close to the leader, the drag force dominates over thrust, causing the follower to decelerate and move further behind the leader. Conversely, when the distance between swimmers is larger than the steady-state spacing, the thrust force overcomes drag causing the follower to accelerate and the pair to move closer; see, e.g., top right of Figures~\ref{fig:transient}(a) and (b), respectively. For either heaving or pitching, the thrust and drag forces on the follower are balanced on average after steady state has been reached, effectively leading to zero acceleration and reach constant separation distance between the swimmers on average. }
%%%%%%%%%%%%%%%%%%%%%%%%%%%

\begin{figure*}
\centering
\includegraphics[scale=1]{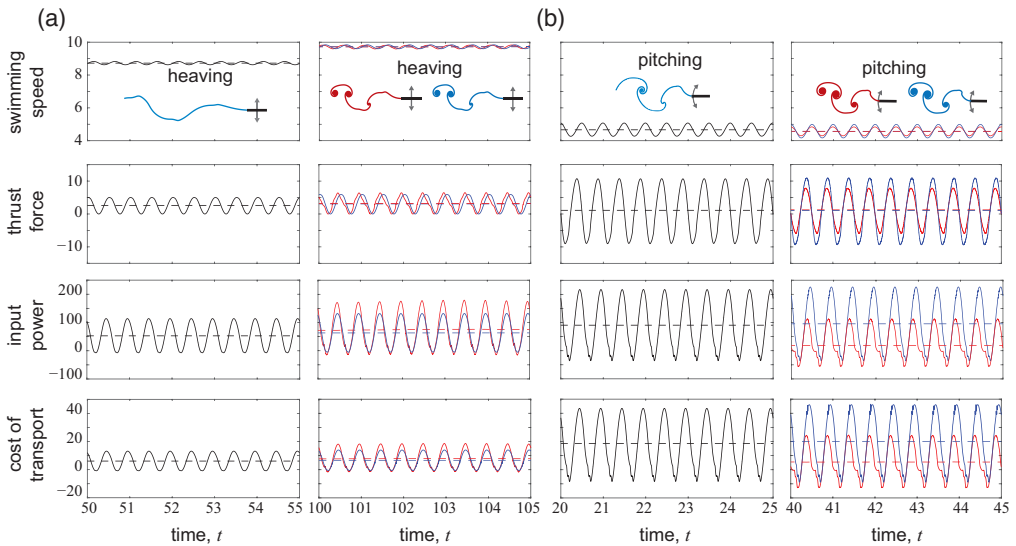}
\caption[]{\footnotesize
Instantaneous swimming performance (time-dependent speed, thrust, input power, and cost of transport versus time) for a single and pair of swimmers undergoing (a) heaving at $A_h = 0.3$ and (b) pitching at $A_p = 15^{\circ}$, respectively. 
%Instantaneous swimming properties of a single swimmer and a pair of swimmers undergoing flapping motions: (a) heaving at $A_h = 0.3$ and (b) pitching at $A_p = 15^{\circ}$. 
Results are shown after the swimmers have reached steady state. From top to bottom, the swimming speed, thrust force, input power and cost of transport are shown. Solid lines represent the instantaneous values  and dashed lines represent time-period averages.}
\label{fig:example}
\end{figure*}

We next evaluate the advantages of these  formations in terms of the speed and energetics of the pair of swimmers in comparison to swimming alone. Figure~\ref{fig:example} shows details of the time evolution at steady state of a single and pair of swimmers 
for the first relative equilibrium $d_\h/\lambda\approx 1$ and $d_\p/\lambda \approx 1$ shown in Figure~\ref{fig:stability}, where hydrodynamic interactions are strongest. From top to bottom, we report the swimming speed, thrust force, input power and cost of transport versus time. Instantaneous values are shown in solid lines and period-average values in dashed lines. For the heaving motion, the average speed of the pair is about 10\% higher than the speed of the single swimmer, consistent with experimental observations on heaving foils \citep{Ramananarivo2016}. However, the input power required to maintain these heaving motions in the presence of hydrodynamic interactions is also higher (about 30\%). Consequently, the cost of transport of the heaving pair is about 20\% higher than a single heaving swimmer. These results suggest that heaving swimmers can enhance their speed by swimming in a pair. However, this enhancement in swimming speed is achieved at an energetic cost. 

For pitching swimmers, the speed of the formation is comparable to that of the single swimmer (about 2\% slower). However, the follower's input power is significantly reduced (about 70\% less than the single pitching swimmer). This reduction in input power is due to the hydrodynamic benefits highlighted in Figure~\ref{fig:flow_field}(b).
Correspondingly, the cost of transport of the pair of pithing swimmers drops by 30\% compared to swimming alone. 

\begin{figure*}
\centering
\includegraphics[scale=1]{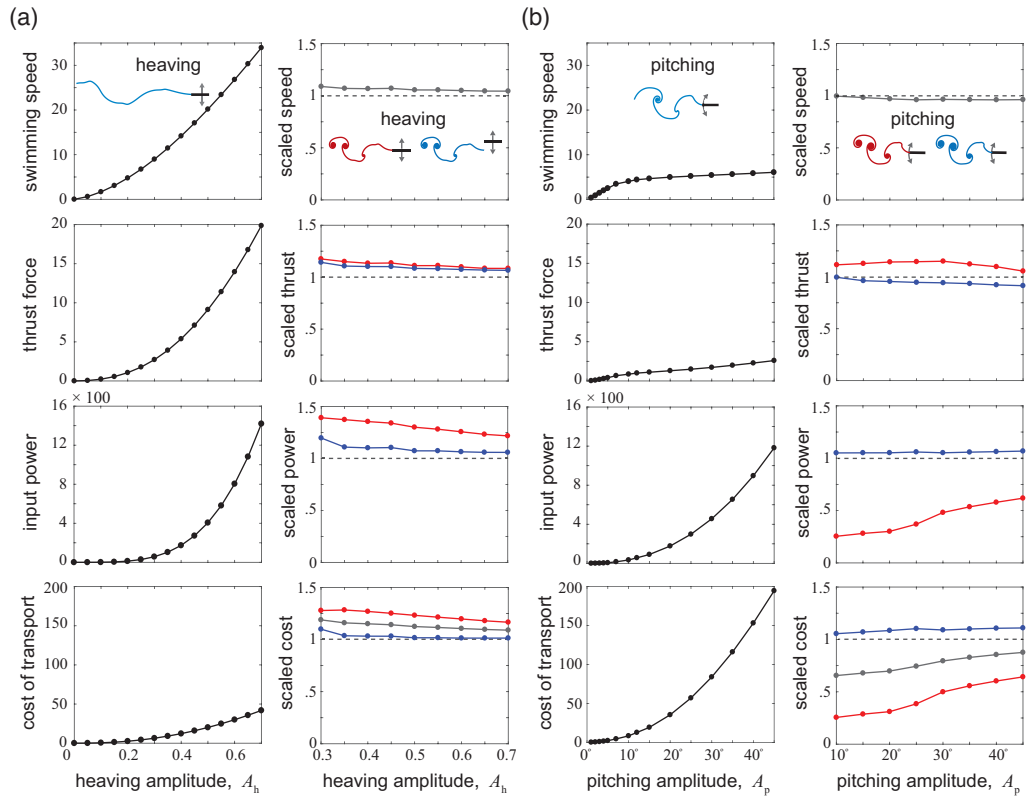}
\caption[]{\footnotesize Swimming performance (average speed, thrust, input power, and cost of transport) versus flapping amplitude for a single and pair of swimmers undergoing (a) heaving and (b) pitching motions, respectively.  From top to bottom, average values of the swimming speed, thrust force, input power and cost of transport. Left columns (black markers) in (a) and (b) show the results for single swimmers. For the pair of swimmers, all of the results are scaled by the corresponding quantity values for a single swimmer. The blue and red markers represent the results for the follower and leader, respectively. The grey markers are the school average.
 }
\label{fig:amplitude}
\end{figure*}

Figure~\ref{fig:amplitude} explores the effect of the flapping amplitude on the period-average values of the swimming speed, thrust force, input power, and cost of transport, after the swimmers have reached steady state. Specifically, we examine the range $A_\h \in [0, 0.7]$ and $A_p \in [0^{\circ}, 45^{\circ}]$ for single swimmers and 
$A_\h \in [0.3, 0.7]$ and $A_p \in [10^{\circ}, 45^{\circ}]$ for pairs of swimmers, where 
small amplitudes are ignored to ensure that hydrodynamic interactions are sufficient for the spontaneous emergence of order formations. In pairwise interactions, we report all period-average values normalized by the corresponding values for a single swimmer. 

When swimming alone, whether by heaving or pitching, an increase in the flapping amplitude monotonically increases the swimming speed, thrust, input power and cost of transport; see left columns of Figure~\ref{fig:amplitude}(a) and (b). Here, the swimming speed versus flapping amplitude for single swimmers is a reproduction of the results in Figure~\ref{fig:compare}(a,b).

Across all heaving amplitudes, the pairwise formation is about 5-10\% faster than that of a single heaving swimmer.  Both the leader and follower experience an increase in thrust compared to the single swimmer, but require more power to swim in formation compared to swimming alone, with extra power demand on the follower. The cost of transport of the heaving formation is thus slightly higher (around $15\%$) compared to swimming alone.  Thus, heaving swimmers slightly enhance their swimming speed when in formation, albeit at a higher cost of transport.

%----
%----------
%----
%----------

The formation of pitching swimmers is about 5\% slower than swimming alone for almost all flapping amplitudes. The leader experiences 
consistently lower thrust and the follower  consistently higher thrust compared to swimming alone. However, while the power demand on the leader is comparable to the single swimmer, the power demand on the follower is significantly reduced for all amplitudes. Taken together, these results lead to slightly higher cost of transport for the leader and significantly lower cost of transport for the follower compared to swimming alone. Indeed, the cost of transport of the follower is a fraction of the single swimmer (around $25\%$ at best), which in turn, causes the formation to save a significant amount of power (around $35\%$ at best) compared to swimming alone. These results imply that although the pairwise formation of pitching swimmers experiences no enhancement in swimming speed compared to swimming alone, it reduces the cost of transport by a significant amount.

{To gain additional insights into the information contained in the wake of the leader and the hydrodynamic mechanisms that mediate the reduction in power consumption and stability of the pairwise formation,  we propose a reduced-order model based on the flow field induced by a single swimmer. 
%First, we assess the hypothesis that the input power reduction in the pitching swimmers is mediated by the wake-induced flow of the leader being favorable to the follower's flapping motion.
We compute the flow field generated behind a single heaving or pitching swimmer, and we consider a virtual ``point'' follower placed at any location $(x_o,y_o)$ in the swimmer's wake and undergoing lateral oscillations $y(t) = y_o + A \sin(2\pi t)$, where $A$ is the oscillation amplitude. We set $A$ to $A_{\rm h}$ in the wake of a heaving swimmer and $A_{\rm p}$ in the wake of a pitching swimmer. In either case, the wake is blind to the existence of the virtual follower. We ask whether there are particular locations in the swimmer's wake that are favorable to the follower's flapping motion. 
To address this question, we define a flow agreement parameter $\mathbb{Z}(x_o,y_o)$ that quantifies the agreement between the flow velocity in the wake of the single swimmer  and the prescribed oscillations of the virtual follower,
%--
\begin{equation}
\textrm{flow agreement parameter:} \quad  \mathbb{Z} =  \dfrac{1}{T} \int_{t_s} ^{t_s+T} \dot{y}(x_o,y_o,t) v(x_o,y_o,t) dt,
\label{eq:flow_agreement}
\end{equation}
%--
where $t_s$ is an arbitrary time after steady state has been reached, $T$ is the flapping period, $\dot{y}(x_o,y_o,t)$ is the lateral velocity of the follower, and $v(x_o,y_o,t)$ is the $y$-component of the flow velocity evaluated at the follower's location. 
%We choose the vertical component of the follower's velocity, since in both heaving and pitching the plates prescribed displacement are mostly in $y$ direction. 
Positive values of the flow agreement parameter imply a beneficial interaction between the flow and the follower's flapping motion, whereas negative values indicate a detrimental one.} %\ek{is $y$ computed at the leading edge for the heaving swimmers, and at the trailing edge for the pitching swimmers?}

\begin{figure*}
\centering
\includegraphics[scale=1]{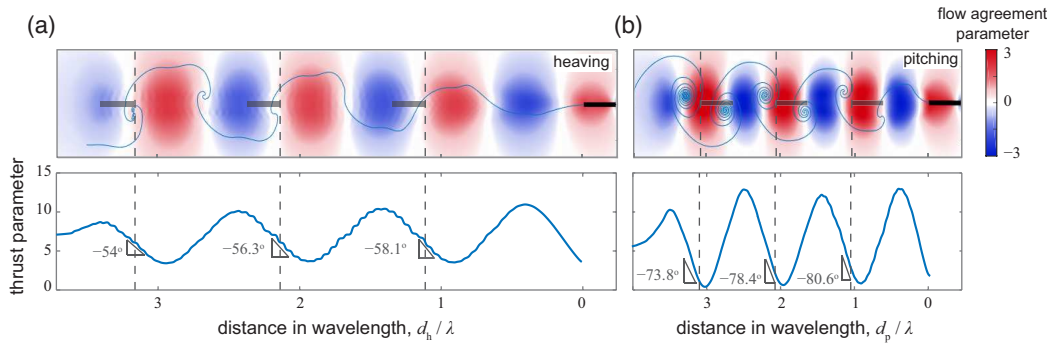}
\caption[]{\footnotesize {Flow agreement parameter and thrust parameter experienced by a hypothetical point follower undergoing prescribed oscillations in the wake of a single swimmer that does not see the follower. Top row shows flow agreement parameter field in the wake of (a) heaving and (b) pitching swimmers. The grey plates represent the steady-state position of the followers in the first, second and third stable spacings found from solving the system with pairwise interactions (Figure~\ref{fig:stability}). In both cases, the distance of the regions with maximum flow agreement from the leading plate is very close to integer multiples of the wavelength ($d_{\h, \p}/\lambda = 1, 2,3$). Bottom row shows thrust parameter as a function of distance. The dashed lines represent the head and tail positions of the heaving and pitching follower, respectively. The negative slopes of plot at the steady-state distances imply linear stability of the follower to in-line perturbations.
 The prescribed amplitudes are $A_\h = 0.3$ and $A_\p = 15^\circ$. }
 }
\label{fig:flow_agreement}
\end{figure*}

{The flow agreement parameter in the wake of a single heaving and pitching swimmer is shown in the top row of Figure~\ref{fig:flow_agreement}. The hypothetical follower is undergoing the same oscillatory motion everywhere in the wake of a single heaving or pitching swimmer at respective amplitudes $A_\h = 0.3$ and $A_\p = 15^\circ$. Red regions  indicate where the flow velocity in the swimmer's wake and the hypothetical follower's motion agree. Interestingly, regions of maximum flow agreement are located at almost integer multiples of the wavelength $\lambda = U/f$ of the single swimmer, similarly to the locations of the stable equilibria in fully-coupled pairwise formations. Superimposed onto Figure~\ref{fig:flow_agreement}(a, c), we show a snapshot of the free vortex sheet of the swimmer, as well as the location of the actual follower at steady state obtained from our pairwise interacting swimmers. As noted previously, in the heaving case, the leading edge of the follower is located close to the integer multiples of $\lambda$, while in pitching, the follower's trailing edge is located at integer multiples of $\lambda$. Interestingly, in the heaving case, the leading edge of the follower is located at the intersection of the red and blue regions of the flow agreement parameter, that is at the location where the flow agreement parameter transitions from favorable to unfavorable. For the pitching swimmer, the follower is mostly located within the red region where the flow agreement parameter is favorable. This effectively means that a higher surface area of the pitching follower experiences a flow field favorable to its motion, whereas part of the heaving follower undergoes negative flow agreement. This mechanism could be responsible for the increased efficiency of the pitching formation in comparison to the heaving formation.} 

%{This is true in both heaving and pitching cases, however, the different equilibrium spacing of heaving and pitching followers result in their different power consumptions. Namely, the leading edge of the heaving follower is positioned close to the integer multiples of the wavelength, while for the pitching follower the trailing edge is placed at these points. This effectively means that a higher surface area of the pitching follower experiences a flow field favorable to its motion, whereas part of the heaving follower undergoes negative flow agreement. }

%----
%----------

{We next examine the stability of pairwise formation in the context of the simpler model based on the wake of a single swimmer and a hypothetical follower. We specifically consider the case where the virtual follower is positioned in-line behind the single swimmer.
It is well-established that the thrust of a self-propelled flapping swimmer scales with the square of the swimmer's lateral velocity relative to the surrounding fluid's velocity \citep{Triantafyllou1993, Floryan2017, Newbolt2019}. We thus define the thrust parameter}
{
%--
\begin{equation}
\textrm{Thrust parameter:} \quad \mathbb{X}=  \dfrac{1}{T} \int_{t_s} ^{t_s+T}(v - \dot{y} )^2dt.
\label{eq:flow_agreement}
\end{equation}
%-- 
%where  $v$ is the vertical velocity of the leader-induced flow and $\dot{y}$ is the vertical velocity of the follower's leading and trailing edge in heaving and pitching plates, respectively. 
The thrust parameter acts as a measure of the period-average thrust but not an exact value of thrust.
We plot the thrust parameter as a function of the follower's downstream location of heaving and pitching swimmers in the bottom row of Figure~\ref{fig:flow_agreement}. The thrust parameter is minimum at three locations where the flow agreement parameter is maximum. This is due to the fact that higher agreement between the follower's oscillation and the flow implies smaller difference in the follower's lateral speed relative to the flow and therefore smaller thrust. 
Superimposed onto these plots are the three equilibria  at steady-state obtained from our pairwise simulations in Figure~\ref{fig:stability} (vertical dashed lines). We next argue that the slope of the thrust parameter at these locations is an indicator of the stability of the pairwise formation. To this end, recall that at steady state, the thrust $F_x(x_o,t)$ and skin drag $D(x_o, t)$ balance each other on average, and the follower experiences zero net acceleration.  Namely, $ \langle F_x(x_o) \rangle -   \langle D(x_o)  \rangle = 0$, where the time average notation $\langle (\cdot) \rangle = (1/T) \int_{t_s}^{t_s+T} (\cdot)dt$ is introduced for brevity. If we perturb the horizontal position of the follower by $\delta x$, since skin drag depends only  on the relative fluid's velocity tangential to the plate, it is reasonable to assume that its change due to in-line positional perturbations is negligible $ \langle D(x_o + \delta x)  \rangle \approx \langle D(x_o)  \rangle$.  We thus arrive at the period-average equation
$\langle F_x(x_o + \delta x)\rangle - \langle F(x_o) \rangle = m \langle \delta \ddot{x}\rangle$.
This equation provides a condition for the linear stability of the pairwise formation in the context of the  (single swimmer / virtual follower) model: if the slope of the period-average thrust relative to the horizontal position is negative, the system is linearly stable to perturbations in the horizontal position. Otherwise, the perturbation grows and the pair leaves their relative spacing at steady state. Since the thrust parameter $\mathbb{X}$ is an approximation of period-average thrust, it suffices to obtain the slope of  $\mathbb{X}$ with respect to $\delta x$ to gauge the stability of the formation. The slope is negative at the steady state positions in both heaving and pitching swimmers (bottom row of Figure~\ref{fig:flow_agreement}). Further, the slope of these locations decreases as the distance between the two swimmers increases. This is consistent with Figure~\ref{fig:stability} where the third stable position was found to be less stable than the second and the second slightly less stable than the first. Finally, the significantly higher slope of the thrust parameter in pitching compared to heaving is consistent with the observations in Figure~\ref{fig:stability}, where pitching formations were found to be more stable.
}

%\textcolor{blue}{The flow agreement parameter is based on the wake of a single swimmer and doesn't require computations of the full pairwise interactions (that can be computationally expensive in most instances). With prior knowledge of the follower's spacing relative to the leader, this parameter provides us with predictions of the required input power and sheds light on the hydrodynamic mechanism underlying pairwise interactions. 

%%%%%%%%%%%%%%%%%%%%%%%%

\section{Conclusion}

%\textcolor{red}{will probably have to edit this paragraph depending on where the comparison paragraph ends up}
We analyzed the locomotion dynamics of actively flapping swimmers interacting passively via the fluid medium in the context of the vortex sheet model. Within the two-swimmer model, we showed that hydrodynamic interactions lead to stable ordered formations, in which the follower falls into specific positions in the wake of the leader, and the pair travels together at the same speed. This well-ordered `schooling' behavior occurs for both heaving and pitching swimmers. Group cohesion is tighter and more stable for pitching swimmers.
In heaving, the school swims slightly faster compared to swimming alone, about 5-10\% faster, albeit at a similar increases in cost of transport, especially for the follower (about 20\% higher cost for the follower and 15\% for the formation). When pitching, the school swims at  a slightly (about 5\%) lower speed but has significant energetic benefits, with up to 35\% reduction in cost of transport for the formation and up to 75\% for the follower. 
{Detailed comparison of our findings with previously known results are in order. Physical experiments and numerical simulations report stable pairwise formations in hydrodynamically-interacting swimmers.  
The experiments of \citep{Ramananarivo2016} using pairs of purely heaving rigid foils in tandem found that the foils stabilize at particular discrete gap distances, and that these formations were usually accompanied by an increase in the swimming speed of the pair ($10-20\%$ compared to swimmer alone). The increase in speed was observed up to three wavelengths away from the leader, however, its effect quickly diminished with distance. 
Numerical simulations of pairs of interacting flapping swimmers provided more details on swimming energetics. \cite{Zhu2014} used an immersed boundary method to study the dynamics of two flexible filaments undergoing heaving oscillations at their leading edges at Reynolds number $= 200$. They reported an increase in both the swimming speed and input power of the pair compared to swimming alone. These changes were only reported for pairs in \textit{compact} configurations. In this configuration, the leading edge of the follower is almost touching the trailing edge of the leader and the narrow space between them causes the pair to behave like one long filament. The increased speed and power requirements seemed to completely disappear for pairs in  \textit{regular} configurations characterized by an increased distance between the swimmers and velocity and power equal to a single swimmer. \cite{Dai2018} studied the swimming dynamics of multiple flexible filaments under combined pitching and heaving motions at the leading edge. However, the heaving motion's amplitude was much smaller than the tail's displacement due to pitching. For two filaments swimming in tandem, they reported a decrease of about $18\%$ in the cost of transport when the swimmers were in compact configurations. The regular configurations was found to be energetically beneficial, but only by about $2-3\%$ compared to swimming alone. \cite{Park2018} also found a decrease of about $15\%$ in power for a pair of flexible filaments, when swimming close to one another. The increase in speed relative to swimming alone was found to be negligible.}

{We examined pairwise interactions of purely heaving and pitching rigid swimmers, thus isolating heaving from pitching as opposed to the studies of flexible heaving filament that combine both effects. We found that for each flapping mode, the swimmers reach stable steady-state formations with constant distances. The flapping mode had a significant impact on the stability and swimming energetics of the pair. We observed a slight increase in the swimming speed of the heaving pair (up to $10\%$) at the expense of higher cost of transport. For pitching swimmers, the swimming speed was not affected much by the pairwise interaction, but we found a significant decrease in the input power of the follower (up to $70\%$ for small amplitudes). 
In contrast to the findings of \cite{Zhu2014, Park2018}, where the effects of the pairwise interactions quickly vanished with increasing distance, our vortex sheet model observed these effects at longer distances, up to three swimming wavelengths, consistent the experiments of \cite{Ramananarivo2016}. The discrepancy is most likely due to the relatively small Reynolds number in \cite{Zhu2014} ($\mathrm{Re} = 200$), causing the wake-induced flow to diffuse faster due to higher viscous forces. \cite{Ramananarivo2016} reported a much larger Reynolds number ($\mathrm{Re} = 10^3 - 10^4$) in their experimental set-up. The higher Reynolds numbers in the experiments are consistent with our inviscid model. At this inviscid regime, the flow inertia is dominant, causing the wake of the leader to live longer in the fluid. In this regime, the effects of hydrodynamic interaction on stability and energetics decreased with distance, but much more gradually.
 }

{In sum, our results are consistent with numerical and experimental findings of heaving foils~\citep{Zhu2014,Park2018, Peng2018,Lin2020,Becker2015, Ramananarivo2016, Newbolt2019}, but go beyond these results in two major ways. 
Firstly, we completely separated the flapping modes, heaving and pitching,
probed the effect of each one on the stability, speed, and energetic performance of the school, and we showed that the flapping mode affects the tightness and stability of the formation, as well as the cost of transport in school compared to swimming alone. Secondly, we analyzed these formations in the context of a simpler model consisting of the wake of a single swimmer and a hypothetical point follower. We defined an empirical flow agreement parameter and showed that regions where the wake-induced flow and the follower's periodic motion agree are consistent with the stable formations observed in pairwise interactions of heaving and pitching swimmers. The reduced-order model also highlights that the heaving mode is less favorable energetically because, in steady state formations of heaving swimmers, the follower is positioned such that it experiences negative agreement with the ambient flow. We also employed the simpler model to make predictions about the stability of the pairwise formation, consistent with our findings that the pitching mode leads to tighter and more stable formation. 
Indeed, an alternative interpretation of our results is that they reveal how active changes in the flapping mode can be used to control, via hydrodynamic interactions, the school emergent properties, including the school speed, energetics, and cohesion. For example, to save energy or quickly overcome large perturbations, swimmers can adopt a pitching mode.  }

These findings could be instrumental for understanding the role of the fluid medium as a mediator of the physical interactions between swimming fish, and to assess the hydrodynamic benefits to fish schooling.  Fish have more complex flapping motions than simple heaving and pitching~\citep{Ayancik2020, Van2019, Lin2019}, and the compliance of the fish body is believed to play an important role in the flapping efficiency and ability to extract energy from ambient flows \citep{Beal2006, Lucas2014, Jusufi2017, Quinn2014}. These considerations, as well as extensions to arrays of swimmers in-tandem and side-by-side, potentially flapping at different amplitudes and phases as in~\cite{Newbolt2019}, will be treated in future works.

\paragraph{Acknowledgment.} The authors would like to thank Michael J. Shelley and Leif Ristroph for interesting conversations. This work is partially supported by the National Science Foundation grant CBET 15-12192, the Office of Naval Research grants 12707602 and N00014-17-1-2062, and the Army Research Office grant W911NF-16-1-0074. 
%E.K. acknowledges partial support from the Simons Foundation for a sabbatical period in 2016-2017.

\bibliographystyle{jfm}
\bibliography{references}

\begin{thebibliography}{62}
\expandafter\ifx\csname natexlab\endcsname\relax\def\natexlab#1{#1}\fi
\def\au#1{#1} \def\ed#1{#1} \def\yr#1{#1}\def\at#1{#1}\def\jt#1{\textit{#1}}
  \def\bt#1{#1}\def\bvol#1{\textbf{#1}} \def\vol#1{#1} \def\pg#1{#1}
  \def\publ#1{#1}\def\arxiv#1{#1}\def\org#1{#1}\def\st#1{\textit{#1}}

\bibitem[Abrahams \& Colgan(1985)]{Abrahams1985}
{\sc \au{Abrahams, M.~V.} \& \au{Colgan, P.~W.}} \yr{1985}  \at{Risk of
  predation, hydrodynamic efficiency and their influence on school structure}.
  \jt{Environmental Biology of Fishes}  \bvol{13}~(3),  \pg{195--202}.

\bibitem[Alben(2009)]{Alben2009}
{\sc \au{Alben, S.}} \yr{2009}  \at{Wake-mediated synchronization and drafting
  in coupled flags}.  \jt{Journal of Fluid Mechanics}  \bvol{641},  \pg{489}.

\bibitem[Alben \& Shelley(2008)]{Alben2008}
{\sc \au{Alben, S.} \& \au{Shelley, M.~J.}} \yr{2008}  \at{Flapping states of a
  flag in an inviscid fluid: bistability and the transition to chaos}.
  \jt{Physical review letters}  \bvol{100}~(7),  \pg{074301}.

\bibitem[Ayancik {\em et~al.\/}(2020)Ayancik, Fish \& Moored]{Ayancik2020}
{\sc \au{Ayancik, F.}, \au{Fish, F.~E} \& \au{Moored, K.~W.}} \yr{2020}
  \at{Three-dimensional scaling laws of cetacean propulsion characterize the
  hydrodynamic interplay of flukes' shape and kinematics}.  \jt{Journal of the
  Royal Society Interface}  \bvol{17}~(163),  \pg{20190655}.

\bibitem[Beal {\em et~al.\/}(2006)Beal, Hover, Triantafyllou, Liao \&
  Lauder]{Beal2006}
{\sc \au{Beal, D.~N.}, \au{Hover, F.~S.}, \au{Triantafyllou, M.~S.}, \au{Liao,
  J.~C.} \& \au{Lauder, G.~V.}} \yr{2006}  \at{Passive propulsion in vortex
  wakes}.  \jt{Journal of Fluid Mechanics}  \bvol{549},  \pg{385}.

\bibitem[Becker {\em et~al.\/}(2015)Becker, Masoud, Newbolt, Shelley \&
  Ristroph]{Becker2015}
{\sc \au{Becker, A.~D.}, \au{Masoud, H.}, \au{Newbolt, J.~W.}, \au{Shelley,
  M.~J.} \& \au{Ristroph, L.}} \yr{2015}  \at{{Hydrodynamic schooling of
  flapping swimmers}}.  \jt{Nature Communications}  \bvol{6},  \pg{8514}.

\bibitem[Blondeaux {\em et~al.\/}(2005)Blondeaux, Fornarelli, Guglielmini,
  Triantafyllou \& Verzicco]{Blondeaux2005}
{\sc \au{Blondeaux, P.}, \au{Fornarelli, F.}, \au{Guglielmini, L.},
  \au{Triantafyllou, M.~S.} \& \au{Verzicco, R.}} \yr{2005}  \at{Numerical
  experiments on flapping foils mimicking fish-like locomotion}.  \jt{Physics
  of Fluids}  \bvol{17}~(11),  \pg{113601}.

\bibitem[Borazjani(2008)]{Borazjani2008}
{\sc \au{Borazjani, I.}} \yr{2008} {\em Numerical simulations of
  fluid-structure interaction problems in biological flows\/}.
  \publ{University of Minnesota}.

\bibitem[Buchholz \& Smits(2008)]{Buchholz2008}
{\sc \au{Buchholz, James~HJ} \& \au{Smits, Alexander~J}} \yr{2008}  \at{The
  wake structure and thrust performance of a rigid low-aspect-ratio pitching
  panel}.  \jt{Journal of fluid mechanics}  \bvol{603},  \pg{331}.

\bibitem[Dabiri(2009)]{Dabiri2009}
{\sc \au{Dabiri, John~O}} \yr{2009}  \at{Optimal vortex formation as a unifying
  principle in biological propulsion}.  \jt{Annual review of fluid mechanics}
  \bvol{41},  \pg{17--33}.

\bibitem[Dai {\em et~al.\/}(2018)Dai, He, Zhang \& Zhang]{Dai2018}
{\sc \au{Dai, Longzhen}, \au{He, Guowei}, \au{Zhang, Xiang} \& \au{Zhang,
  Xing}} \yr{2018}  \at{Stable formations of self-propelled fish-like swimmers
  induced by hydrodynamic interactions}.  \jt{Journal of The Royal Society
  Interface}  \bvol{15}~(147),  \pg{20180490}.

\bibitem[Dong {\em et~al.\/}(2006)Dong, Mittal \& Najjar]{Dong2006}
{\sc \au{Dong, H}, \au{Mittal, R} \& \au{Najjar, FM}} \yr{2006}  \at{Wake
  topology and hydrodynamic performance of low-aspect-ratio flapping foils}.
  \jt{Journal of Fluid Mechanics}  \bvol{566},  \pg{309}.

\bibitem[Eldredge(2019)]{Eldredge2019}
{\sc \au{Eldredge, J.~D.}} \yr{2019} {\em Mathematical Modeling of Unsteady
  Inviscid Flows\/}.  \publ{Springer}.

\bibitem[Fang(2016)]{Fang2016}
{\sc \au{Fang, F.}} \yr{2016}  \at{Hydrodynamic interactions between
  self-propelled flapping wings}. PhD thesis, New York University.

\bibitem[Filella {\em et~al.\/}(2018)Filella, Nadal, Sire, Kanso \&
  Eloy]{Filella2018}
{\sc \au{Filella, A.}, \au{Nadal, F.}, \au{Sire, C.}, \au{Kanso, E.} \&
  \au{Eloy, C.}} \yr{2018}  \at{Model of collective fish behavior with
  hydrodynamic interactions}.  \jt{Physical review letters}  \bvol{120}~(19),
  \pg{198101}.

\bibitem[Floryan {\em et~al.\/}(2017)Floryan, Van~Buren, Rowley \&
  Smits]{Floryan2017}
{\sc \au{Floryan, Daniel}, \au{Van~Buren, Tyler}, \au{Rowley, Clarence~W} \&
  \au{Smits, Alexander~J}} \yr{2017}  \at{Scaling the propulsive performance of
  heaving and pitching foils}.  \jt{Journal of Fluid Mechanics}  \bvol{822},
  \pg{386--397}.

\bibitem[Franck \& Breuer(2017)]{Franck2017}
{\sc \au{Franck, Jennifer~A} \& \au{Breuer, Kenneth~S}} \yr{2017}  \at{Unsteady
  high-lift mechanisms from heaving flat plate simulations}.  \jt{International
  Journal of Heat and Fluid Flow}  \bvol{67},  \pg{230--239}.

\bibitem[Garrick {\em et~al.\/}(1937)]{Garrick1937}
{\sc \au{Garrick, IE} \& \au{others}} \yr{1937}  \at{Propulsion of a flapping
  and oscillating airfoil}.  \jt{NACA report}  \bvol{567},  \pg{419--427}.

\bibitem[Gazzola {\em et~al.\/}(2016)Gazzola, Tchieu, Alexeev, de~Brauer \&
  Koumoutsakos]{Gazzola2016}
{\sc \au{Gazzola, M.}, \au{Tchieu, A.~A.}, \au{Alexeev, D.}, \au{de~Brauer, A.}
  \& \au{Koumoutsakos, P.}} \yr{2016}  \at{Learning to school in the presence
  of hydrodynamic interactions}.  \jt{Journal of Fluid Mechanics}  \bvol{789},
  \pg{726--749}.

\bibitem[Huang {\em et~al.\/}(2016)Huang, Nitsche \& Kanso]{Huang2016}
{\sc \au{Huang, Y.}, \au{Nitsche, M.} \& \au{Kanso, E.}} \yr{2016}
  \at{Hovering in oscillatory flows}.  \jt{Journal of Fluid Mechanics}
  \bvol{804},  \pg{531--549}.

\bibitem[Huang {\em et~al.\/}(2018)Huang, Ristroph, Luhar \& Kanso]{Huang2018}
{\sc \au{Huang, Y.}, \au{Ristroph, L.}, \au{Luhar, M.} \& \au{Kanso, E.}}
  \yr{2018}  \at{Bistability in the rotational motion of rigid and flexible
  flyers}.  \jt{Journal of Fluid Mechanics}  \bvol{849},  \pg{1043--1067}.

\bibitem[Jones(2003)]{Jones2003}
{\sc \au{Jones, M.~A.}} \yr{2003}  \at{The separated flow of an inviscid fluid
  around a moving flat plate}.  \jt{Journal of Fluid Mechanics}  \bvol{496},
  \pg{405}.

\bibitem[Jones(2005)]{Jones2005}
{\sc \au{Jones, M. A .and~Shelley, M.~J.}} \yr{2005}  \at{Falling cards}.
  \jt{Journal of Fluid Mechanics}  \bvol{540},  \pg{393--425}.

\bibitem[Jusufi {\em et~al.\/}(2017)Jusufi, Vogt, Wood \& Lauder]{Jusufi2017}
{\sc \au{Jusufi, A.}, \au{Vogt, D.~M.}, \au{Wood, R.~J.} \& \au{Lauder, G.~V.}}
  \yr{2017}  \at{Undulatory swimming performance and body stiffness modulation
  in a soft robotic fish-inspired physical model}.  \jt{Soft robotics}
  \bvol{4}~(3),  \pg{202--210}.

\bibitem[Kanso \& Newton(2009)]{Kanso2009}
{\sc \au{Kanso, E.} \& \au{Newton, P.~K.}} \yr{2009}  \at{Passive locomotion
  via normal-mode coupling in a submerged spring-mass system}.  \jt{Journal of
  Fluid Mechanics}  \bvol{641},  \pg{205}.

\bibitem[Kanso \& Tsang(2014)]{Kanso2014}
{\sc \au{Kanso, E.} \& \au{Tsang, A. C.~H.}} \yr{2014}  \at{Dipole models of
  self-propelled bodies}.  \jt{Fluid Dynamics Research}  \bvol{46}~(6),
  \pg{061407}.

\bibitem[Kanso \& Tsang(2015)]{Kanso2015}
{\sc \au{Kanso, E} \& \au{Tsang, A C~H}} \yr{2015}  \at{Pursuit and
  synchronization in hydrodynamic dipoles}.  \jt{Journal of Nonlinear Science}
  \bvol{25(5)},  \pg{1141}.

\bibitem[Lauder {\em et~al.\/}(2011)Lauder, Lim, Shelton, Witt, Anderson \&
  Tangorra]{Lauder2011}
{\sc \au{Lauder, G.~V.}, \au{Lim, J.}, \au{Shelton, R.}, \au{Witt, C.},
  \au{Anderson, E.} \& \au{Tangorra, J.~L.}} \yr{2011}  \at{Robotic models for
  studying undulatory locomotion in fishes}.  \jt{Marine Technology Society
  Journal}  \bvol{45}~(4),  \pg{41--55}.

\bibitem[Liao(2007)]{Liao2007}
{\sc \au{Liao, J.~C.}} \yr{2007}  \at{A review of fish swimming mechanics and
  behaviour in altered flows}.  \jt{Philosophical Transactions of the Royal
  Society B: Biological Sciences}  \bvol{362}~(1487),  \pg{1973--1993}.

\bibitem[Liao {\em et~al.\/}(2003)Liao, Beal, Lauder \&
  Triantafyllou]{Liao2003}
{\sc \au{Liao, J.~C.}, \au{Beal, D.~N.}, \au{Lauder, G.~V.} \&
  \au{Triantafyllou, M.~S.}} \yr{2003}  \at{Fish exploiting vortices decrease
  muscle activity}.  \jt{Science}  \bvol{302}~(5650),  \pg{1566--1569}.

\bibitem[Lin {\em et~al.\/}(2019)Lin, Wu \& Zhang]{Lin2019}
{\sc \au{Lin, X.}, \au{Wu, J.} \& \au{Zhang, T.}} \yr{2019}  \at{Performance
  investigation of a self-propelled foil with combined oscillating motion in
  stationary fluid}.  \jt{Ocean Engineering}  \bvol{175},  \pg{33--49}.

\bibitem[Lin {\em et~al.\/}(2020)Lin, Wu, Zhang \& Yang]{Lin2020}
{\sc \au{Lin, Xingjian}, \au{Wu, Jie}, \au{Zhang, Tongwei} \& \au{Yang,
  Liming}} \yr{2020}  \at{Self-organization of multiple self-propelling
  flapping foils: energy saving and increased speed}.  \jt{Journal of Fluid
  Mechanics}  \bvol{884}.

\bibitem[Lucas {\em et~al.\/}(2014)Lucas, Johnson, Beaulieu, Cathcart, Tirrell,
  Colin, Gemmell, Dabiri \& Costello]{Lucas2014}
{\sc \au{Lucas, K.~N.}, \au{Johnson, N.}, \au{Beaulieu, W.~T.}, \au{Cathcart,
  E.}, \au{Tirrell, G.}, \au{Colin, S.~P.}, \au{Gemmell, B.~J.}, \au{Dabiri,
  J.~O.} \& \au{Costello, J.~H.}} \yr{2014}  \at{Bending rules for animal
  propulsion}.  \jt{Nature communications}  \bvol{5}~(1),  \pg{1--7}.

\bibitem[Marras {\em et~al.\/}(2015)Marras, Killen, Lindstr{\"o}m, McKenzie,
  Steffensen \& Domenici]{Marras2015}
{\sc \au{Marras, S.}, \au{Killen, S.~S.}, \au{Lindstr{\"o}m, J.}, \au{McKenzie,
  D.~J.}, \au{Steffensen, J.~F.} \& \au{Domenici, P.}} \yr{2015}  \at{Fish
  swimming in schools save energy regardless of their spatial position}.
  \jt{Behavioral ecology and sociobiology}  \bvol{69}~(2),  \pg{219--226}.

\bibitem[Moored \& Quinn(2019)]{Moored2019}
{\sc \au{Moored, K.~W.} \& \au{Quinn, D.~B.}} \yr{2019}  \at{Inviscid scaling
  laws of a self-propelled pitching airfoil}.  \jt{AIAA Journal}
  \bvol{57}~(9),  \pg{3686--3700}.

\bibitem[Newbolt {\em et~al.\/}(2019)Newbolt, Zhang \& Ristroph]{Newbolt2019}
{\sc \au{Newbolt, J.~W.}, \au{Zhang, J.} \& \au{Ristroph, L.}} \yr{2019}
  \at{{Flow interactions between uncoordinated flapping swimmers give rise to
  group cohesion}}.  \jt{Proceedings of the National Academy of Sciences}
  \bvol{116},  \pg{201816098}.

\bibitem[Nitsche \& Krasny(1994)]{Nitsche1994}
{\sc \au{Nitsche, M.} \& \au{Krasny, R.}} \yr{1994}  \at{A numerical study of
  vortex ring formation at the edge of a circular tube}.  \jt{Journal of Fluid
  Mechanics}  \bvol{276},  \pg{139--161}.

\bibitem[Oza {\em et~al.\/}(2019)Oza, Ristroph \& Shelley]{Oza2019}
{\sc \au{Oza, A.~U.}, \au{Ristroph, L.} \& \au{Shelley, M.~J.}} \yr{2019}
  \at{Lattices of hydrodynamically interacting flapping swimmers}.
  \jt{Physical Review X}  \bvol{9}~(4),  \pg{041024}.

\bibitem[Park \& Sung(2018)]{Park2018}
{\sc \au{Park, Sung~Goon} \& \au{Sung, Hyung~Jin}} \yr{2018}  \at{Hydrodynamics
  of flexible fins propelled in tandem, diagonal, triangular and diamond
  configurations}.  \jt{Journal of Fluid Mechanics}  \bvol{840},  \pg{154}.

\bibitem[Partridge(1982)]{Partridge1982}
{\sc \au{Partridge, B.~L.}} \yr{1982}  \at{The structure and function of fish
  schools}.  \jt{Scientific american}  \bvol{246}~(6),  \pg{114--123}.

\bibitem[Partridge \& Pitcher(1979)]{Partridge1979}
{\sc \au{Partridge, B.~L.} \& \au{Pitcher, T.~J.}} \yr{1979}  \at{Evidence
  against a hydrodynamic function for fish schools}.  \jt{Nature}
  \bvol{279}~(5712),  \pg{418--419}.

\bibitem[Peng {\em et~al.\/}(2018)Peng, Huang \& Xi-Yun]{Peng2018}
{\sc \au{Peng, Ze-Rui}, \au{Huang, Haibo} \& \au{Xi-Yun, Lu}} \yr{2018}
  \at{Collective locomotion of two closely spaced self-propelled flapping
  plates}.  \jt{Journal of Fluid Mechanics}  \bvol{849},  \pg{1068--1095}.

\bibitem[Quinn {\em et~al.\/}(2014)Quinn, Lauder \& Smits]{Quinn2014}
{\sc \au{Quinn, D~B}, \au{Lauder, G~V} \& \au{Smits, A~J}} \yr{2014}
  \at{Scaling the propulsive performance of heaving flexible panels}.
  \jt{Journal of fluid mechanics}  \bvol{738},  \pg{250}.

\bibitem[Ramananarivo {\em et~al.\/}(2016)Ramananarivo, Fang, Oza, Zhang \&
  Ristroph]{Ramananarivo2016}
{\sc \au{Ramananarivo, S.}, \au{Fang, F.}, \au{Oza, A.}, \au{Zhang, J.} \&
  \au{Ristroph, L.}} \yr{2016}  \at{Flow interactions lead to orderly
  formations of flapping wings in forward flight}.  \jt{Phys. Rev. Fluids}
  \bvol{1},  \pg{071201}.

\bibitem[Saffman(1992)]{Saffman1992}
{\sc \au{Saffman, P.~G.}} \yr{1992} {\em Vortex dynamics\/}.  \publ{Cambridge
  university press}.

\bibitem[Shaw(1978)]{Shaw1978}
{\sc \au{Shaw, E.}} \yr{1978}  \at{Schooling fishes: the school, a truly
  egalitarian form of organization in which all members of the group are alike
  in influence, offers substantial benefits to its participants}.  \jt{American
  Scientist}  \bvol{66}~(2),  \pg{166--175}.

\bibitem[Sheng {\em et~al.\/}(2012)Sheng, Ysasi, Kolomenskiy, Kanso, Nitsche \&
  Schneider]{Sheng2012}
{\sc \au{Sheng, J.~X.}, \au{Ysasi, A.}, \au{Kolomenskiy, D.}, \au{Kanso, E.},
  \au{Nitsche, M.} \& \au{Schneider, K.}} \yr{2012}  \at{Simulating vortex
  wakes of flapping plates}.  \bt{In {\em Natural locomotion in fluids and on
  surfaces\/}},  \pg{pp. 255--262}.  \publ{Springer}.

\bibitem[Smits(2019)]{Smits2019}
{\sc \au{Smits, Alexander~J}} \yr{2019}  \at{Undulatory and oscillatory
  swimming}.  \jt{Journal of Fluid Mechanics}  \bvol{874}.

\bibitem[Taneda(1965)]{Taneda1965}
{\sc \au{Taneda, S.}} \yr{1965}  \at{Experimental investigation of vortex
  streets}.  \jt{Journal of the Physical Society of Japan}  \bvol{20}~(9),
  \pg{1714--1721}.

\bibitem[Tchieu {\em et~al.\/}(2012)Tchieu, Kanso \& Newton]{Tchieu2012}
{\sc \au{Tchieu, A.~A.}, \au{Kanso, E.} \& \au{Newton, P.~K.}} \yr{2012}
  \at{The finite-dipole dynamical system}.  \jt{Proceedings of the Royal
  Society A: Mathematical, Physical and Engineering Sciences}
  \bvol{468}~(2146),  \pg{3006--3026}.

\bibitem[Triantafyllou {\em et~al.\/}(1993)Triantafyllou, Triantafyllou \&
  Grosenbaugh]{Triantafyllou1993}
{\sc \au{Triantafyllou, George~S}, \au{Triantafyllou, MS} \& \au{Grosenbaugh,
  MA}} \yr{1993}  \at{Optimal thrust development in oscillating foils with
  application to fish propulsion}.  \jt{Journal of Fluids and Structures}
  \bvol{7}~(2),  \pg{205--224}.

\bibitem[Triantafyllou {\em et~al.\/}(2000)Triantafyllou, Triantafyllou \&
  Yue]{Triantafyllou2000}
{\sc \au{Triantafyllou, M.~S.}, \au{Triantafyllou, G.~S.} \& \au{Yue, D.~K.}}
  \yr{2000}  \at{Hydrodynamics of fishlike swimming}.  \jt{Annual review of
  fluid mechanics}  \bvol{32}~(1),  \pg{33--53}.

\bibitem[Tsang \& Kanso(2013)]{Tsang2013}
{\sc \au{Tsang, A. C.~H.} \& \au{Kanso, E.}} \yr{2013}  \at{Dipole interactions
  in doubly periodic domains}.  \jt{Journal of Nonlinear Science}
  \bvol{23}~(6),  \pg{971--991}.

\bibitem[Van~Buren {\em et~al.\/}(2019)Van~Buren, Floryan \& Smits]{Van2019}
{\sc \au{Van~Buren, T.}, \au{Floryan, D.} \& \au{Smits, A.~J.}} \yr{2019}
  \at{Scaling and performance of simultaneously heaving and pitching foils}.
  \jt{AIAA Journal}  \bvol{57}~(9),  \pg{3666--3677}.

\bibitem[Verma {\em et~al.\/}(2018)Verma, Novati \& Koumoutsakos]{Verma2018}
{\sc \au{Verma, S.}, \au{Novati, G.} \& \au{Koumoutsakos, P.}} \yr{2018}
  \at{Efficient collective swimming by harnessing vortices through deep
  reinforcement learning}.  \jt{Proceedings of the National Academy of
  Sciences}  \bvol{115}~(23),  \pg{5849--5854}.

\bibitem[Weihs(1973)]{Weihs1973}
{\sc \au{Weihs, D.}} \yr{1973}  \at{{Hydromechanics of fish schooling}}.
  \jt{Nature}  \bvol{241},  \pg{241290a0}.

\bibitem[Weihs(1975)]{Weihs1975}
{\sc \au{Weihs, D.}} \yr{1975}  \at{Some hydrodynamical aspects of fish
  schooling}.  \bt{In {\em Swimming and flying in nature\/}},  \pg{pp.
  703--718}.  \publ{Springer}.

\bibitem[Wen \& Lauder(2013)]{Wen2013}
{\sc \au{Wen, L.} \& \au{Lauder, G.}} \yr{2013}  \at{Understanding undulatory
  locomotion in fishes using an inertia-compensated flapping foil robotic
  device}.  \jt{Bioinspiration \& biomimetics}  \bvol{8}~(4),  \pg{046013}.

\bibitem[White(1979)]{White1979}
{\sc \au{White, F.~M.}} \yr{1979} {\em Fluid mechanics\/}.  \publ{Tata
  McGraw-Hill Education}.

\bibitem[Wolfgang {\em et~al.\/}(1999)Wolfgang, Anderson, Grosenbaugh, Yue \&
  Triantafyllou]{Wolfgang1999}
{\sc \au{Wolfgang, M.~J.}, \au{Anderson, J.~M.}, \au{Grosenbaugh, M.~A.},
  \au{Yue, D.~K.} \& \au{Triantafyllou, M.~S.}} \yr{1999}  \at{Near-body flow
  dynamics in swimming fish}.  \jt{Journal of Experimental Biology}
  \bvol{202}~(17),  \pg{2303--2327}.

\bibitem[Wu(1971)]{Wu1971}
{\sc \au{Wu, T.}} \yr{1971}  \at{Hydromechanics of swimming propulsion. part 1.
  swimming of a two-dimensional flexible plate at variable forward speeds in an
  inviscid fluid}.  \jt{Journal of Fluid Mechanics}  \bvol{46}~(2),
  \pg{337--355}.

\bibitem[Zhu {\em et~al.\/}(2014)Zhu, He \& Zhang]{Zhu2014}
{\sc \au{Zhu, Xiaojue}, \au{He, Guowei} \& \au{Zhang, Xing}} \yr{2014}
  \at{Flow-mediated interactions between two self-propelled flapping filaments
  in tandem configuration}.  \jt{Physical review letters}  \bvol{113}~(23),
  \pg{238105}.

\end{thebibliography}

%\newpage

%\newpage

\appendix
\section{Vortex sheet model}
\label{sec:appendix}

The coupled fluid-structure interaction between the swimming plate and the surrounding fluid  is simulated using an inviscid vortex sheet model. 
%The resulting fluid motion induces hydrodynamic forces and moment $F_x$, $F_y$ and $M$ on the plate. 
In viscous fluids, boundary layer vorticity is formed along the sides of the swimmer, and it is swept away at the swimmer's tail to form a shear layer that rolls up into vortices. 
In the vortex sheet model, the swimmer is approximated by a bound vortex sheet, denoted by $l_\b$, whose strength ensures that no fluid flows through the rigid plate, and the separated shear layer is
approximated by a free regularized vortex sheet $l_\w$ at the trailing edge of the swimmer.
The total shed circulation $\Gamma$  in the vortex sheet is determined so as to satisfy the Kutta condition at the trailing edge, which is given 
in terms of the tangential velocity components above and below the 
bound sheet and ensures that the pressure jump across the sheet vanishes at the trailing edge. 

To express these concepts mathematically, it is convenient to use the complex notation $z = {\rm x} + \ii {\rm y}$, where $\ii = \sqrt{-1}$ and $({\rm x},{\rm y})$ denote the components
of an arbitrary point in the plane. 
The bound vortex sheet $l_\b$ is described by its position $z_\b(s,t)$ and strength $\gamma(s,t)$, 
where $s\in[-l,l]$ denotes the arc length along the sheet $l_\b$.
The separated sheet $l_\w$ is described by its position $z_\w(\Gamma,t)$, $\Gamma\in[0,\Gamma_\w]$ where $\Gamma$ is the Lagrangian circulation around the portion of the separated sheet between its free end in the spiral center and the point $z_\w(\Gamma,t)$. 
The parameter $\Gamma$ defines the vortex sheet strength $\gamma=d\Gamma/ds$.

By linearity of the problem, the complex velocity $w(z,t) = u(z,t) - \mathrm{i}v(z,t)$ 
is a superposition of the contributions due to the  bound and free vortex sheets 
%----
\begin{equation}
w(z,t)= w_\b(z,t) + w_\w(z,t).
\label{flowfield}
\end{equation}
%-----
In practice, the free sheet $l_\w$ is regularized
using the vortex blob method to prevent the growth of the Kelvin-Helmholtz 
instability. The bound sheet $l_\b$ is not regularized in order to preserve the invertibility of the map between the sheet strength and the normal velocity along the sheet. The velocity components $w_\b(z,t)$ and $w_\w(z,t)$  induced by the bound and free vortex sheets, respectively, are given by
%---
\begin{equation}
\begin{aligned}
w_\b(z,t) = 
\int_{-l}^{l} K_o(z-z_{b}(s,t))\gamma(s,t)\, ds, \quad
w_\w(z,t) =  
\int_0^{\Gamma_\w}K_{\delta}(z-z_\w(\Gamma,t))\, d\Gamma,
\end{aligned}
\label{fluidvelo}
\end{equation}
where $K_\delta$ is the vortex blob kernel, with regularization parameter $\delta$,
\begin{equation}
K_\delta (z) = \frac{1}{2\pi \mathrm{i}}\frac{\overline{z}}{|z|^2+\delta^2}, \qquad \overline{z} = x - \ii y
\end{equation}
If $z$ is a point on the bound sheet for which $\delta = 0$, $w_\b$ is to be computed in the principal value sense.  

%----------
\begin{figure*}
\label{fig:appendix}
\centering
\includegraphics[scale=1]{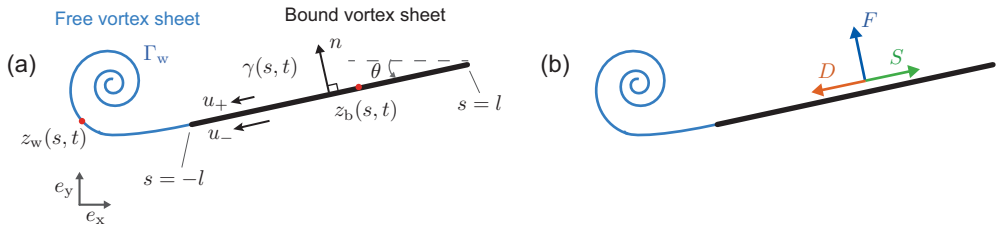}
\caption[]{\footnotesize (a) Schematic of the vortex sheet model for a two-dimensional flapping swimmer. (b) Depiction of the different hydrodynamic forces acting on the swimmer. 
}
\label{fig:}
\end{figure*}
%----------

The position of the bound vortex sheet $z_\b$
is determined from the plate's flapping $(y(t),\theta(t))$ and swimming $x(t)$ motions.
The corresponding sheet strength $\gamma(s,t)$
is determined by imposing the no penetration boundary condition 
on the plate, together with conservation of total circulation. Let $n(s,t)=-\sin\theta+\ii \cos\theta$ be the upward normal to the plate, 
the no penetration boundary condition is given by
%--------
\begin{equation}
\Real \left[ w n\right]_{z_b}= 
\Real \left[  w_{\rm swimmer} n\right],
\end{equation}
%-----
where 
%-------
\begin{equation}
w_{\rm swimmer} =  \dot{x} - \ii \dot{y}-\mathrm{i}\dot{\theta}\left[\bar{z}_b - ({x}- \ii {y})\right].
\end{equation}
%--------
Conservation of the fluid circulation implies that
$\int_{l_b} \gamma(s,t) ds + \Gamma_\w (t) = 0$.

The circulation parameter $\Gamma$ along the free vortex sheet $z_\w(\Gamma,t)$
is determined by the circulation shedding rates $\dot\Gamma_\w$, according to 
the Kutta condition, which states that the fluid velocity at the trailing edge is finite and tangent to the flyer.
The Kutta condition can be obtained from the Euler equations by enforcing that, at the trailing edge, the difference in pressure 
across the swimmer is zero. To this end, we integrate the balance of momentum equation for inviscid planar flow
along a closed contour containing the vortex sheet and trailing edge,
%---
\begin{equation}
[p]_\mp(s) = p_-(s) - p_+(s) =  - \frac{d\Gamma(s,t)}{dt} - \frac{1}{2}(u_-^2 - u_+^2),
\label{pressure}
\end{equation}
%%---
where $\Gamma(s,t) = \Gamma_\w+\int_{-l}^{s} \gamma(s',t)ds'$, $-l \le s \le l$, 
is the circulation within the contour and $p_{\mp}(s,t)$ 
and $u_\mp(s,t)$  denote the limiting pressure and tangential 
slip velocities on both sides of the swimmer. 
Since the pressure difference across the free sheet is zero, 
it also vanishes at the trailing edge by continuity, which implies that
%----
\begin{equation}
\dot{\Gamma}_{\w}=-\frac{1}{2}(u_-^2 - u_+^2)|_{s=- l}.
\label{sheddingrates}
\end{equation}
The values of
$u_-$ and $u_+$ 
are obtained from the average tangential velocity component and 
from the velocity jump at the trailing edge, given by the sheet strength, evaluated at $s=- l$
%---
\begin{equation}
\overline{u}={u_+ + u_-\over 2}=\Imag[(w-w_{\rm swimmer})n]~,
\qquad 
u_--u_+= \gamma .
\label{veloedge}
\end{equation}
%---
Once shed, the vorticity in the free sheet moves with the flow. 
Thus the 
parameter $\Gamma$ assigned to each particle $z_\w(\Gamma,t)$
is the value of $\Gamma_\w$ at the instant it is shed from the trailing edge.
The evolution of the free vortex sheet $z_\w$ is obtained by 
advecting it in time with the fluid velocity,
%---
\begin{equation}
\dot{\bar{z}}_{\w} = w_\w(z_{\w},t) +w_\b(z_{\w},t).
\label{freesheets}
\end{equation}
%---

\section{Forces and moments}
\label{app:forces}
The hydrodynamic force acting on the swimmer due to the pressure difference across the swimmer is given by,
%---
\begin{equation}
\begin{split}
 \int_{l_b} n[p]_\mp d s = -F \sin\theta  + \ii F \cos\theta , \qquad
\end{split}
\label{eq:force}
\end{equation}
where $F = \int_{l_b} [p]_\mp d s$. 
The hydrodynamic moment acting on the swimmer about its leading edge is given by
%---
\begin{equation}
M = \Real\left[ \int_{l_b}\mathrm{i}\overline{n}(z_{\rm le}-z_b)[p]_\mp ds \right],
\label{eq:moment}
\end{equation}
%--
where $z_{\rm le}$ is position of the leading edge $s = \pm l$.

%\section{Leading Edge Suction}
%\label{sec:le}

It is known that the strength of the bound vortex sheet exhibits an inverse square root singularity at the edges (\cite{Saffman1992, Eldredge2019}). The singularity at the trailing edge is regularized by enforcing the Kutta condition as discussed above. To regularize the singularity at the leading edge, we introduce a force parallel to the plate known as leading edge suction \citep{Eldredge2019}. 
%At the leading edge the velocity is infinite, which means pressure is negatively infinite and its integral over the vanishingly small region surrounding the leading edge results in a finite suction directed parallel to the plate. 
Following the derivation provided in \citet{Eldredge2019}, we write the suction force, in dimensionless form as
%---
\begin{equation}
\label{eq:suction_force}
S =  2 \pi  e^{\textrm{i} \theta} \sigma^2,
\end{equation}
%---
where $\sigma$ is the suction parameter defined as 
\begin{equation}
\label{eq:suction_param}
%\sigma = \frac{1}{2}(\dot{y_b} - V_{\infty}) + \sum_{j=1}^{N_v} \frac{\Gamma_j}{2\pi l} \textrm{Re} \Big(  \frac{ \tilde{z_j} + l}{ \tilde{z_j} - l} \Big) ^{1/2},
\sigma = \frac{1}{2}(\dot{y} - l\dot{\theta} \cos \theta ) + \int_{l_b} \frac{\gamma (s, t)}{2 \pi l} \textrm{Re} \Big(  \frac{ \tilde{z}(s, t) + l}{ \tilde{z}(s, t) - l} \Big) ^{1/2} ds
\end{equation}
%
%\textcolor{blue}{first term is the vertical velocity of the center of the plate. in our notation, $y(t)$ is the position of the leading edge} 
where $\tilde{z}(s,t) = z(s,t) - z e^{\textrm{i} \theta}$, is the complex position of any vortex sheet present in the fluid written in the plate's frame of reference. $\dot{y} - l\dot{\theta} \cos \theta$ is the velocity of the center of the plate in the $y$-direction. Note that in equation~\ref{eq:suction_force}, the suction force is always positive (always a thrust force) and parallel to the plate.

Note that the majority of the suction force is due to the vertical motion of the leading edge relative to the surrounding fluid. For the pitching swimmer, since the leading edge has no vertical motion, the contribution of the leading edge suction force to the total thrust force of the swimmer is negligible. This is confirmed by our numerical experiments on a single pitching swimmer.

%\section{Skin Drag}
%\label{sec:sd}

Last, we introduce a drag force $D$ that emulates the effect of skin friction due to fluid viscosity. This force is based on the Blasius laminar boundary layer theory as implemented by \citet{Fang2016} in the context of the vortex sheet model. Blasius theory provides an empirical formula for skin friction on one side of a horizontal plate of length $2l$ placed in fluid of density $\rho_f$ and uniform velocity $U$.
In dimensional form, Blasius formula is $D =  \frac{1}{2} \rho_\f (2l) (c_\f)U^2$, where the skin friction coefficient $C_\f = {0.664}/{\sqrt{\rm Re}}$ is given in terms of the Reynolds number $\textrm{Re} = {\rho_f U (2l)}/{\mu}$. 
Substituting back in the empirical formula leads to $D =  C_d U^{3/2}$, where $C_d = 0.664 \sqrt{ \rho_f \mu (2l)}$.
Following \cite{Fang2016}, we write a modified expression of the drag force for a swimming plate
\begin{equation}
D =  C_d (\overline{U}_+ ^{3/2} + \overline{U}_- ^{3/2}),
%0.664 \sqrt{ \rho_f \mu (2l)} (\overline{U}_+ ^{3/2} + \overline{U}_- ^{3/2}),
\end{equation}
where $\overline{U}_{\pm}$ are the spatially-averaged tangential fluid velocities on the upper and lower side of the plate, respectively, relative to the swimming velocity $U$,
\begin{equation}
\overline{U}_{\pm}(t) = \frac{1}{2l} \int_{-l} ^{l} u_{\pm} (s,t)ds - U.
\end{equation}
%
%Then the drag force due to skin friction is given by
%%---
%\begin{equation}
%F_d =  C_d (\overline{U}_+ ^{3/2} + \overline{U}_- ^{3/2}),
%\end{equation}
%%--
%where the drag coefficient $C_d$ is now defined as
% %
%\begin{equation}
%C_d = 0.664 \sqrt{2 \rho \mu l}.
%\end{equation}
%%
We estimate $C_d$ to be approximately $0.02$ in the experiments of \citet{Ramananarivo2016}.
%Since in our problem the plates are placed in a fluid with no background flow ($U_{\infty}=0$), we arrive at the following expression for the drag force experienced by each plate
%%---
%\begin{equation}
%D =   \dfrac{C_d}{2l}  \int_{-l} ^{l} \big( u_{-}(s,t) + u_{+} (s,t) \big) ds. 
%\label{eq:drag}
%\end{equation}

\section{Numerical implementation}
\label{sec:numerics}

The bound vortex sheet is discretized by $2n+1$ 
point vortices at $z^b(t)$ with strength $\Delta\Gamma=\gamma\Delta s$.
These vortices are located at Chebyshev 
points that cluster at the two ends of the swimmer.
Their strength is determined by enforcing no penetration at the 
midpoints between the vortices, together with conservation of circulation.
The free vortex sheet is discretized by regularized point vortices at $z^{\w}(t)$, that is released from the trailing edge at each timestep with circulation given by \eqref{sheddingrates}.
The free point vortices move with the discretized fluid velocity 
while the bound vortices move with the swimmer's velocity. 
The discretization of 
equations (\ref{eq:eom}) and (\ref{sheddingrates}, \ref{freesheets}) 
yields 
a coupled system of ordinary differential
evolution equations for the
swimmer's position, the shed circulation, and the free vorticity,
that is integrated in time using the 4th order 
Runge-Kutta scheme. 
%All implementation details are given next.
The details of the shedding algorithm are given in \citet{Nitsche1994}. 
%which we follow closely here.
The numerical values of the timestep $\Delta t$, the number of bound vortices $n$,
 and the regularization parameter $\delta$ are chosen 
so that the solution changes little under further refinement.

Finally, to emulate the effect of viscosity, we allow 
the shed vortex sheets to decay gradually by
dissipating each incremental point vortex after 
a finite time $T_{\rm diss}$ from the time it is shed into the fluid.
Larger $T_{\rm diss}$ implies that the vortices stay in the fluid 
for longer times, mimicking the effect of lower fluid viscosity. 
For the results depicted in this study, we used $T_{\rm diss} \in [1.5, 3.5]$ flapping period.
We refer the reader to \cite{Huang2018} for a detailed analysis of the effect of dissipation time on the hydrodynamic forces on a stationary and moving plate in the vortex sheet model. Details of the numerical validation in comparison to \cite{Jones2003} and \cite{Jones2005} are provided in  \cite{Huang2016}.

\section{Swimming Energetics}
\label{sec:efficiency}

Heaving motions are produced by an active heaving force $F_{\rm h}$ acting by the swimmer on the fluid in the $y$-direction. The value of $F_{\rm h}$  is obtained from the balance of linear momentum on the swimmer in the $y$-direction,
%---
\begin{equation}
\begin{split}
\textrm{Heaving:} & \quad  m\ddot{y} = F_y + F_{\rm h} .%\quad \Longrightarrow F_a = m\ddot{y} - F_y
\end{split}
\end{equation}
%---
Here, the hydrodynamic force $F_y$ acting on the swimmer in the $y$-direction is given by \eqref{eq:force}.

Pitching motions are produced by an active moment $M_{\rm p}$ acting by the swimmer on the fluid about the leading edge. The value of $M_{\rm p}$  is obtained from the balance of angular momentum about the swimmer's leading edge,
%---
\begin{equation}
\begin{split}
\textrm{Pitching:} \quad  I \ddot{\theta} - \Imag[m(\dot{x}+\ii\dot{y})w_{\rm l.e.}] & = M + M_{\rm p},
\end{split}
\end{equation}
%---
Here, $I=m(2l)^2/3$ is the swimmer's moment of inertia about the leading edge, $w_{\rm l.e.}$ is the swimmer's velocity at the leading edge, and $M$ is the hydrodynamic moment about the leading edge given in \eqref{eq:moment}.

The power input by the swimmer into the fluid due to heaving and pitching motions, respectively, is given by
%---
\begin{equation}
\begin{split}
\textrm{Heaving:} \quad &P_\h = F_{\rm h} \dot{y}, \\
\textrm{Pitching:} \quad  & P_\p = M_{\rm p} \dot{\theta}.
\end{split}
\end{equation}
%--
Note that in both cases, the leading edge suction and skin drag forces do not contribute to the input power.

%%%%%%

\newpage

\end{document}